# Singular Perturbation of Nonlinear Dynamics by Parasitic Noise

Cheng Li[1,2], Guo-Qiang Wu[2] and Chi-Sang Poon[1]

[1]Harvard-MIT Division of Health Sciences and Technology, Massachusetts Institute of Technology, Cambridge, MA 02139, USA

[2]Department of Mechanics and Engineering Science, Fudan University, Shanghai 200433, People's Republic of China

Correspondence:

    Chi-Sang Poon

    Harvard-MIT Division of Health Sciences and Technology

    Bldg. E25-250

    Massachusetts Institute of Technology

    77 Massachusetts Avenue

    Cambridge, MA 02139

    Tel: +1 617-258-5405; Fax: +1 617-258-7906

    Email: cpoon@mit.edu

**Abstract**

In nonlinear systems analysis, minor fractions of higher-order dynamics are often neglected for simplicity. Here, we show that machine epsilon levels of parasitic higher-order dynamics due to computer roundoff alone can cause divergence of the Hénon attractor to new attractors or instability. The divergence develops exponentially regardless of whether the original or new attractor is chaotic or not. Such singular perturbation by parasitic higher-order dynamics is a novel property of nonlinear dynamics that is of wide practical significance in dynamical systems modeling, simulation and control.

**Introduction**

Chaotic motion is generally characterized by sensitive dependence on initial conditions, a fundamental property first noted by Poincaré and later elaborated by E. Lorenz [1]. In his historic discovery, Lorenz simulated his model equations using second-order integration methods. However, it was soon pointed out that higher-order integration methods with reduced truncation errors could produce different numerical results [2]. A fundamental question posed by S. Smale (as one of the open mathematical challenges for the 21$^{st}$ Century [3]) is whether the Lorenz model indeed supports a chaotic attractor beyond computer simulation; existence of such a geometric attractor was subsequently proved by Tucker [4]. Nevertheless, questions remain as to whether uniform convergence toward this attractor is ever possible in the face of finite truncation errors [5-7] even when quadruple-precision numerical integration with very small step sizes is used [8]. This limitation appears to underlie many chaotic attractors described by differential equations (see [7]). In some cases a supposedly chaotic numerical solution may become spuriously periodic or unstable, or vice versa, depending on the choices of integration methods and step sizes [9-11].

Truncation error represents one type of parasitic noise that may confound nonlinear dynamics computations. Another type of parasitic noise is roundoff (rounding) error, which represents the limit of machine precision (machine epsilon) [12, 13]. Computation of chaotic attractors is susceptible to roundoff errors owing to the property of sensitive dependence on initial conditions. Nevertheless, the resultant numerical orbits may shadow (stay close to) some nearby true chaotic trajectories for a long time provided the errors are sufficiently small and the system is hyperbolic (i.e., the tangent space at almost every point can be split into distinct stable and unstable subspaces) [14-16]. Indeed, the significance of computer roundoff goes beyond the realm of computation of chaotic attractors in that it provides a critical test of the robustness of a mathematical model to system disturbances and/or parameter variations that are bound to occur in real systems.

An important class of parasitic noise not considered in previous studies concerns the provocation of higher-order dynamics. Sizable levels of unmodeled linear or nonlinear dynamics will certainly disturb systems behavior and may even cause instability in control systems [17], but how about trace levels? This is a critical question because minor fractions of higher-order dynamics are often neglected for simplicity in dynamical systems modeling, simulation and control. Here, we show that trace levels of parasitic higher-order dynamics caused by computer roundoff alone could profoundly disrupt chaotic or nonchaotic dynamics with exponential divergence demonstrating singular perturbation, a condition of dramatic changes in systems behavior as some critical parameter tends to certain limiting value [18]. A fundamental implication of this surprising observation is that nonlinear dynamical systems are not only sensitive to initial conditions when they are chaotic, but their behaviors may be highly sensitive to even machine epsilon levels of parasitic higher-order dynamics regardless of whether the system is chaotic or not.

**Results**

To illustrate, we examine the effects of roundoff on the computation of the Hénon map, a two-dimensional discrete-time map derived from some Poincaré section of the Lorenz attractor [19]. Truncation errors are negligible in this case because the Hénon map does not require numerical integration or any series expansion. Since the exact solution cannot be computed, we evaluate the effects of roundoff by comparing two numerical simulations of the Hénon map with slightly different arrangements of some nonlinear terms:

$$\text{Hénon-1a} \qquad \begin{aligned} x_{n+1} &= y_n + 1 - ax_n^2 \\ y_{n+1} &= bx_n \end{aligned} \qquad (1a)$$

$$\text{Hénon-1b} \qquad \begin{aligned} x_{n+1} &= y_n + 1 - (ax_n)x_n \\ y_{n+1} &= bx_n \end{aligned} \qquad (1b)$$

Equations (1a) and (1b) are equivalent mathematically except the nonlinear terms $ax_n^2$ and $(ax_n)x_n$ have different order of multiplication and hence, different roundoff errors. Figure 1A shows that both equations have identical bifurcation maps with well-defined periodic and chaotic regimes that are consistently identified by the largest Lyapunov exponent (LLE) and noise limit (NL), a chaotic measure that is more robust to noise than LLE [20-22]. In the periodic regime (Fig. 1B), the numerical orbits and first return maps for both equations match one another within machine precision (double precision machine epsilon $\sim 10^{-16}$). In the chaotic regime (Fig. 1C), however, the corresponding orbits diverge exponentially to near-maximum separations after ~100 iterations even though they exhibit similar first return maps shadowing the same strange attractor, as proven previously for the Hénon map [16]. The divergence between the two chaotic orbits begins at $n = 1$, when the state variables $x_1$ for both equations first separate by a minute amount $\Delta x_1 = ax_0^2 - (ax_0)x_0 \approx 0$ as a result of differing roundoff errors for the two nonlinear terms. This initial error sets in motion subsequent exponential divergence between the two numerical orbits as determined by their LLE. Subsequent roundoff errors are less significant because the separations $\Delta x_n$, $n > 1$ between the two orbits are now dominated by the increasing divergence resulting from $\Delta x_1$. Thus, computed chaotic orbits are influenced by roundoff mainly through disturbances of initial conditions.

When Eqs. 1a and 1b are recalculated in single precision (machine epsilon ~$10^{-7}$), the resultant bifurcation maps remain unchanged as in Fig. 1A but unlike in Fig. 1C, the corresponding numerical chaotic orbits become identical and do not diverge from one another as they continue to shadow the same strange attractor (not shown). This is because the roundoff errors now become so large that subtle differences in equation form can no longer be resolved and both numerical orbits are rounded to the same values.

Now, consider the following recursive form of the Hénon map:

$$\text{Hénon-2a} \quad \begin{aligned} x_{n+1} &= y_n + 1 - a(z_n + 1 - a(\frac{y_n}{b})(\frac{y_n}{b}))^2 \\ y_{n+1} &= bx_n \\ z_{n+1} &= y_n \end{aligned} \quad (2a)$$

Equation 2a is equivalent to Eq. 1a except that the nonlinear term $x_n^2$ is written recursively as $(z_n + 1 - a(\frac{y_n}{b})(\frac{y_n}{b}))^2$. In theory, the recursive state $z_n = bx_{n-2}$ should be cancelled out after algebraic simplification but roundoff errors leave trace levels of this recursive term in the computed nonlinear term $x_n^2$ giving rise to spurious higher-order dynamics. Surprisingly, Fig. 2A shows that Hénon-2a exhibits a different double-precision bifurcation map than that of Hénon-1a or -1b (Fig. 1A). Not only are the bifurcations to periodic and chaotic regimes shifted and reorganized as indicated by the corresponding values of NL (the LLE cannot be reliably evaluated in this case because the LLE formula [23, 24] for Eq. 2a is ill-conditioned and highly sensitive to roundoff errors), but an apparent equilibrium regime (which actually proves to be a periodic mode with very small amplitude; see Fig. 4D legend) and an unstable regime emerge in Hénon-2a that are not present in Hénon-1a or 1b. The resultant bifurcation maps are dependent on subtle differences in roundoff for the nonlinear term $x_n^2$ and its recursive representation; differing bifurcation maps are obtained for different recursive representations such as in the following (Figs. 2B-2D):

$$\text{Hénon-2b} \qquad x_{n+1} = y_n + 1 - a(z_n + 1 - a(\frac{y_n}{b})^2)^2 \qquad (2b)$$

$$\text{Hénon-2c} \qquad x_{n+1} = y_n + 1 - a(z_n + 1 - a(\frac{y_n}{b})(\frac{y_n}{b}))(z_n + 1 - a(\frac{y_n}{b})(\frac{y_n}{b})) \qquad (2c)$$

$$\text{Hénon-2d} \qquad x_{n+1} = y_n + 1 - a(z_n + 1 - a(\frac{y_n}{b})^2)(z_n + 1 - a(\frac{y_n}{b})^2) \qquad (2d)$$

where the state variables $y_n$, $z_n$ are as defined in Eq. 2a. Equations 2b-2d are all mathematically equivalent to Eq. 2a except for the manner in which the term $x_n^2$ is recursively calculated, yet the resultant bifurcation maps are all different from one another.

When Eqs. 2a-2d are reevaluated in single precision, the resultant bifurcation maps become identical to one another (Fig. 3A) as the roundoff errors are now much larger and are the same for all equations. However, all the single-precision bifurcation maps are again different from their double-precision counterparts (Fig. 2) or those resulting from Eqs. 1a and 1b, which are identical for single- and double-precision computations (Fig. 1A). The profound roundoff-dependent changes in the bifurcation maps for Eqs. 2a-2d (Fig. 2, 3A) indicate singular perturbations by higher-order dynamics, in contrast to the non-singular perturbations in Eqs. 1a-1b for shadowing (Fig. 1A). In Eqs. 2a-2d, the resultant bifurcation maps are dependent on the initial conditions regardless of single- or double-precision computations (not shown).

To gain insight into the singular perturbation of the Hénon map by parasitic higher-order dynamics, we analyze the numerical orbits resulting from Eqs. 1a and 2a at selected values of the bifurcation parameter *a*. In the range roughly 0 < *a* < 0.74 where both Hénon-1a and 2a are in either equilibrium or period-2 oscillation states, both numerical orbits and their first return maps match one another to within machine precision (not shown) similar to Fig. 1B. At higher values of *a*, however, the Hénon-2a orbit diverges exponentially from the original attractor to form totally different attractors regardless of whether the original or the new attractor is chaotic or not (Fig. 4). In two cases, the singularly-perturbed orbit diverges from the original period-2 attractor to form a new oscillatory attractor with higher periodicity (Fig. 4A) or a new chaotic attractor (Fig. 4B). In other cases, the singularly-perturbed orbit diverges from the original chaotic attractor to form another chaotic attractor (Fig. 4C) or a new oscillatory attractor (Fig. 4D), or become unstable (Fig. 4E). In all cases, the singularly-perturbed orbit diverges exponentially from the original attractor and does not shadow it like Hénon-1a and 1b do. The resultant Hénon-2a attractor persists in the steady state even though the effects of roundoff are negligible for $n \gg 1$ compared with the full-blown divergence of the singularly-perturbed orbit. Thus, the initial roundoff at $n = 1$ leads to sustained subsequent divergence to form new attractors under the increasing influence of higher-order dynamics.

**Concluding remarks**

To our knowledge, the foregoing results represent the first demonstration of singular perturbation of nonlinear attractors by parasitic higher-order dynamics to form new attractors. Although a detailed analysis of the theoretical underpinning of this phenomenon is beyond the scope of this Letter, our simulation results suggest that the sensitivity to parasitic higher-order dynamics is due to local instability of the parasitic-induced perturbation. For the range 0 < *a* < 0.74 in the Hénon map, the equilibrium point $\Delta x_n = \Delta y_n = \Delta z_n = 0$ is locally stable for all $n \geq 0$ and the resultant orbits match the original ones within machine precision. For *a* > 0.74 (roughly), $\Delta x_n = \Delta y_n = \Delta z_n = 0$ is locally unstable as a result of higher-order dynamics and the resultant singularly-perturbed orbits are shown to diverge exponentially under the increasing influence of higher-order dynamics to form new attractors, regardless of whether the original or new attractor is chaotic or not. In contrast to the shadowing of chaotic attractors (Fig. 1A) [14-16], sensitivity to parasitic higher-order dynamics is not contingent upon sensitivity to initial conditions. However, the resultant attractors are dependent on the initial conditions as well as the types of parasitic higher-order dynamics and their magnitudes. Unlike the spurious computational chaos/instability/periodicity induced by truncation [9-11] or roundoff errors [5, 25] previously reported, singular perturbation by parasitic higher-order dynamics is a novel property of nonlinear dynamics that is of general applicability to any computational and real-world systems. In particular, the present findings have important implications in the modeling, simulation and control of nonlinear dynamical systems in that even trace levels of unmodeled higher-order dynamics—no matter how miniscule—are not necessarily negligible for simplicity as commonly presumed, as such parasitic noise may be greatly amplified by singular perturbation to eventually disrupt the principal nonlinear dynamics.

**Acknowledgments**

CL was supported by the State Scholarship Fund awarded by the China Scholarship Council. This work was supported in part by U.S. NIH grants HL079503 and RR028241.

**REFERENCES**:

[1] E. N. Lorenz, Journal of the Atmospheric Sciences **20**, 130 (1963).
[2] C. Young, and E. N. Lorenz, J Atmos Sci **23**, 628 (1966).
[3] S. Smale, Math Intell **20**, 7 (1998).
[4] W. Tucker, Found Comput Math **2**, 53 (2002).
[5] R. M. Corless, Comput Math Appl **28**, 107 (1994).
[6] J. Teixeira, C. A. Reynolds, and K. Judd, J Atmos Sci **64**, 175 (2007).
[7] L. Yao, Nonlinear analysis: modeling and control **15**, 109 (2010).
[8] E. N. Lorenz, Tellus **60A**, 806 (2008).
[9] E. N. Lorenz, Physica D **35**, 299 (1989).
[10] E. N. Lorenz, Tellus A **58**, 549 (2006).
[11] R. M. Corless, C. Essex, and M. A. H. Nerenberg, Phys Lett A **157**, 27 (1991).
[12] R. M. Corless, G. W. Frank, and J. G. Monroe, Physica D **46**, 241 (1990).
[13] R. M. Corless, The American Mathematical Monthly **99**, 203 (1992).
[14] D. Anosov, Proc. Steklov. Inst. Math. **90** (1967).
[15] R. Bowen, J Differ Equations **18**, 333 (1975).
[16] S. M. Hammel, J. A. Yorke, and C. Grebogi, B Am Math Soc **19**, 465 (1988).
[17] C. Rohrs *et al.*, in *Proc. 21st 1EEE CDC Conf.*Orlando, Florida, 1982), pp. 3.
[18] D. R. Smith, *Singular-perturbation theory : an introduction with applications* (Cambridge University Press, Cambridge [Cambridgeshire] ; New York, 1985), pp. xvi.
[19] M. Hénon, Communications in Mathematical Physics **50**, 69 (1976).
[20] C.-S. Poon, and M. Barahona, Proceedings of the National Academy of Sciences of the United States of America **98**, 7107 (2001).
[21] M. Barahona, and C.-S. Poon, Nature **381**, 215 (1996).
[22] C.-S. Poon, C. Li, and G.-Q. Wu, arXiv:1004.1427 (http://arxiv.org/abs/1004.1427) (2010).
[23] J. P. Eckmann, and D. Ruelle, Reviews of Modern Physics **57**, 617 (1985).
[24] V. I. Oseledets, Trans. Moscow Math. Soc **19**, 197 (1968).
[25] C. Grebogi, E. Ott, and J. A. Yorke, Phys Rev A **38**, 3688 (1988).

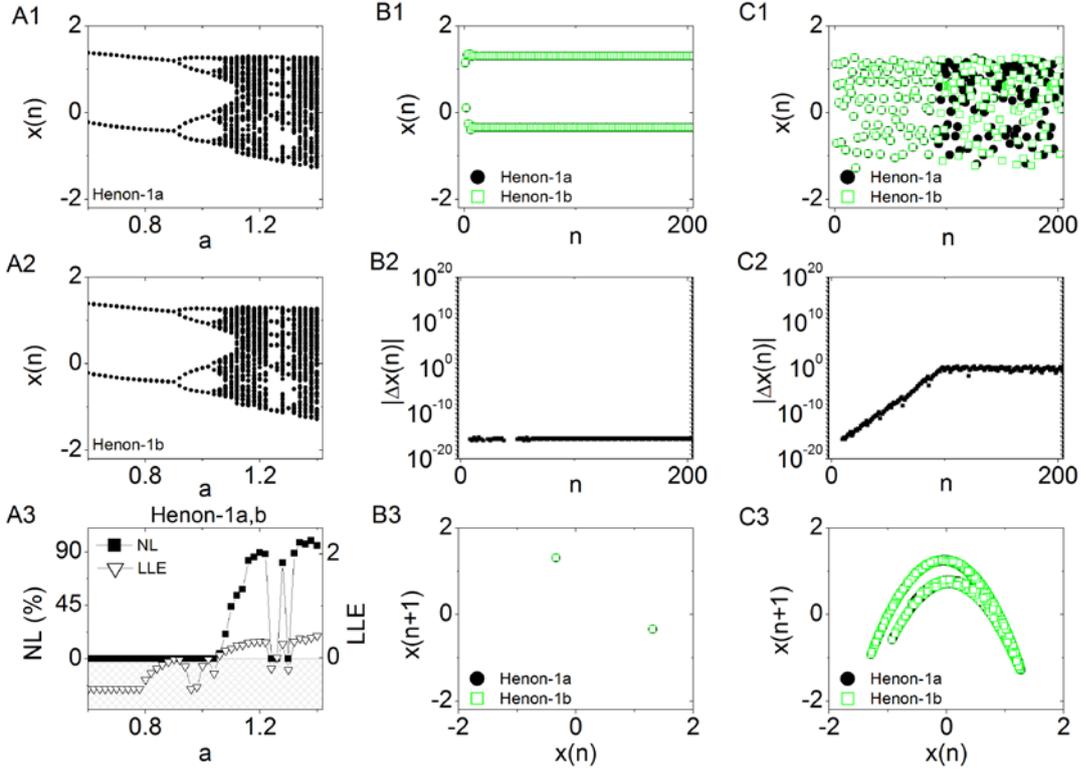

**Fig. 1.** Sensitivity of the Hénon attractor to roundoff errors in initial conditions. Initial conditions for all simulations are Hénon-1a: $x_1=b*0.2+1-a*(0.4)^2$, $y_1=b*0.4$; Hénon-1b: $x_1=b*0.2+1-(a*0.4)*(0.4)$, $y_1=b*0.4$. A1-A3: Identical bifurcation maps of Hénon-1a and 1b (Eqs.1a, 1b) and corresponding NL and LLE results indicating chaotic and nonchaotic regimes. B1-B3: Representative periodic time series from Hénon-1a and 1b (with $a$=0.8, $b$=0.3). From top to bottom are numerical time series, absolute separations between the time series, and corresponding first return maps. C1-C3: Representative chaotic time series from Hénon-1a and 1b ($a$=1.4, $b$=0.3). When Eqs. 1a, 1b are recalculated in single precision, the bifurcation maps remain the same as in (A) but unlike in (C), the corresponding chaotic Hénon-1a and 1b orbits become identical and do not diverge from one another (not shown).

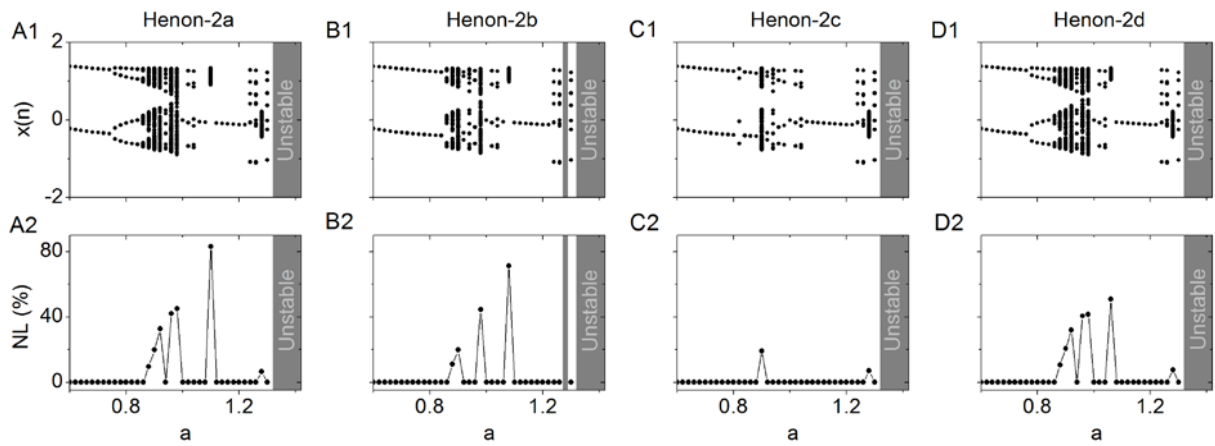

**Fig. 2.** Sensitivity of the Hénon attractor to parasitic higher-order dynamics caused by roundoff errors. A1-D1 are bifurcation maps of 4 different Hénon recursive equations (Eqs. 2a-2d) with initial conditions $x_1=b*0.2+1-a*(0.4)^2$, $y_1=b*0.4$, $z_1=b*0.2$, and $b=0.3$. The resultant singularly-perturbed bifurcation maps are dependent on the initial conditions. A2-D2 are corresponding noise titration results indicating chaotic and nonchaotic regimes.

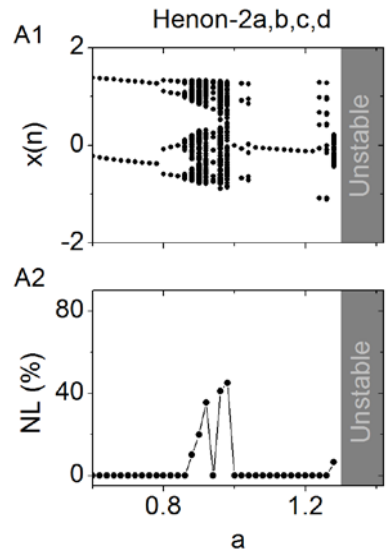

**Fig. 3.** Repeat of Fig. 2 in single-precision simulations. Results are identical for Hénon-2a, b, c and d. The resultant singularly-perturbed bifurcation maps are dependent on the initial conditions.

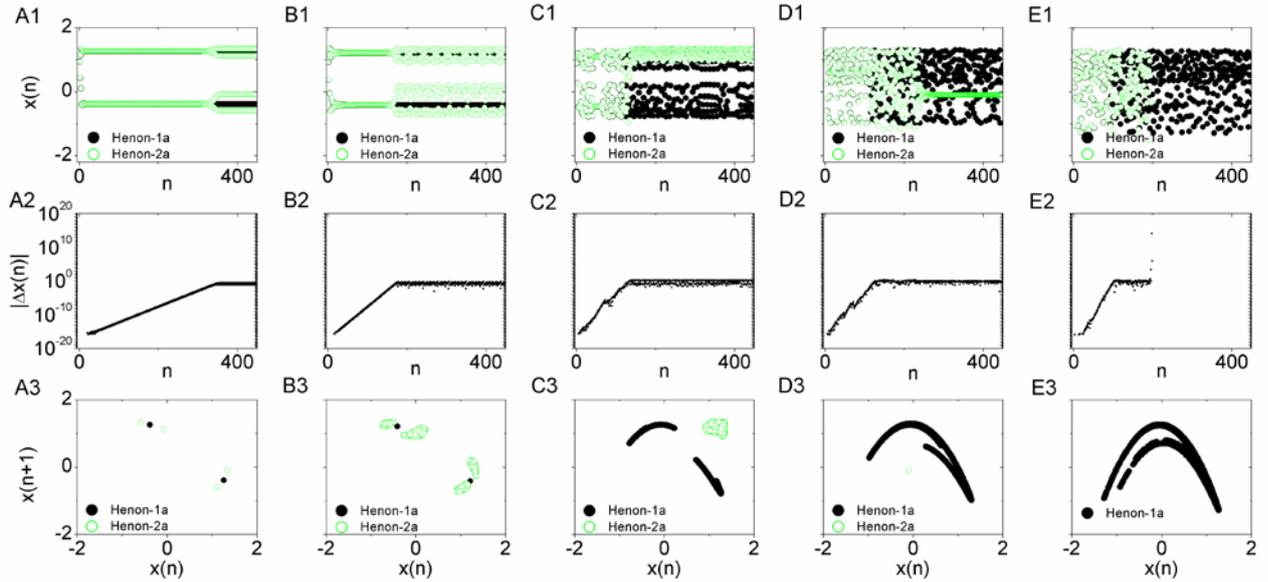

**Fig. 4.** Divergence between Hénon-1a and Hénon-2a. The rows are, from top to bottom: numerical time series, absolute separations between time series and corresponding first return maps. Initial conditions for Hénon-1a and Hénon-2a are as in Figs. 1 and 2. These initial conditions ensure that both equations are initialized the same. (A) $a = 0.80$: period-2 oscillation diverges to apparent period-4 oscillation. (Close examination of the numerical results reveals a subtle period-8 instead.) (B) $a = 0.88$: period-2 oscillation diverges to chaotic fluctuations as indicated by the first return map and by NL. (C) $a = 1.10$: one chaotic mode diverges to another. (D) $a = 1.16$: divergence from chaotic to apparent equilibrium state. (Close examination of the numerical results reveals a subtle period-12 oscillation with very small amplitude instead.) (E) $a = 1.40$: divergence from chaotic to unstable state. For the range roughly $0 < a < 0.74$ (not shown) both numerical orbits demonstrate equilibrium or period-2 oscillation states that match one another within machine precision.